\documentclass[prl,twocolumn,aps,superscriptaddress,showpacs,footinbib,bibnotes]{revtex4-2}
\usepackage{graphicx}
\usepackage{physics,amsmath,amsfonts,amssymb,dsfont,amsthm}
\usepackage{comment}

\usepackage[colorlinks=true,
            citecolor=blue,
            linkcolor=blue,
            urlcolor=blue]{hyperref}
\usepackage{mathtools}
\usepackage{multirow}
\usepackage[utf8]{inputenc}
\theoremstyle{definition}

\usepackage{xcolor}
\begin{document}
\title{
Gottesman-Kitaev-Preskill error-correction with decohered resources}
\author{Rajendra S. Bhati}
\email{rsbhati@cft.edu.pl}
\affiliation{Center for Theoretical Physics, Polish Academy of Sciences, Aleja Lotników 32/46, 02-668 Warszawa, Poland}

\author{T. Bazylewicz} 
\affiliation{Center for Theoretical Physics, Polish Academy of Sciences, Aleja Lotników 32/46, 02-668 Warszawa, Poland}
\affiliation{Faculty of Physics,  University of Warsaw, 02-093 Warsaw, Poland}

\author{Jarosław K.~Korbicz}
\email{jkorbicz@cft.edu.pl}
\affiliation{Center for Theoretical Physics, Polish Academy of Sciences, Aleja Lotników 32/46, 02-668 Warszawa, Poland}

\begin{abstract}

Teleportation-based Gottesman--Kitaev--Preskill (GKP) error correction typically assumes a decoherence-free GKP ancilla Bell pair, an idealization that is difficult to reconcile with the ubiquitous environmental decoherence in bosonic systems. Here, we analyze the performance of the GKP protocol subject to pure-dephasing processes, ubiquitous in realistic situations.
We show that, under this class of interactions, teleportation, instead of projecting the input onto the GKP subspace, transforms GKP error correction into a correlated Pauli--Gaussian noise process that can accumulate over successive correction rounds. Thus, unlike in previously studied noisy GKP scenarios, the code no longer self-corrects. Moreover, repeated noisy GKP teleportation drives the encoded modes out of the logical subspace, progressively degrading the encoded codewords. We show that the resulting leakage rapidly saturates with the number of correction rounds. Finally, we investigate the implications of our results for motional-oscillator-based GKP error-correction platforms.


\end{abstract} 

\maketitle
\raggedbottom

\paragraph*{Introduction.--}
Over the past couple of decades, a variety of bosonic quantum error-correcting (QEC) codes have been proposed and, increasingly, realized experimentally~\cite{CAI202150,Terhal_2020, Mirrahimi_2014, PhysRevX.10.011058, PhysRevX.6.031006, PhysRevA.64.012310}. Among these, the Gottesman-Kitaev-Preskill (GKP) code is particularly attractive because it encodes logical qubits into (non-Gaussian) grid states that are stabilized by Gaussian operations~\cite{PhysRevA.64.012310}. This unique structure enables Clifford quantum computing (QC) and QEC using experimentally accessible Gaussian resources~\cite{PhysRevLett.123.200502,ffln-vd4x}, making the GKP code a leading candidate for achieving fault-tolerant quantum computation (FTQC)~\cite{PRXQuantum.2.020101, Bourassa2021blueprintscalable}.
Approximate GKP states have been experimentally realized in trapped-ion mechanical oscillators~\cite{Fluhmann2019, deNeeve2022, Matsos2025} and superconducting microwave cavities~\cite{Campagne_Ibarcq2020,Eickbusch2022,Sivak2023}. Meanwhile, scalable generation schemes continue to be actively pursued for optical~\cite{Larsen2025,PRXQuantum.3.030301, Takase2023, PhysRevLett.132.230602, Erkılıc2026}, opto-mechanical, and other bosonic platforms~\cite{PhysRevLett.128.170503, 10.1063/5.0197119, PhysRevA.111.022432}. Among these, optical continuous-variable (CV) architectures are particularly attractive for measurement-based quantum computation~\cite{PhysRevLett.97.110501, Asavanant2019} owing to their highly efficient homodyne measurements and deterministic Gaussian entangling operations implemented with linear optics~\cite{Larsen2025,Bourassa2021blueprintscalable}. However, in contrast to other leading platforms, the experimental generation of high-quality optical GKP states with sufficiently large squeezing remains a major challenge and is expected to be one of the dominant limitations to achieving the fault-tolerance threshold~\cite{PhysRevLett.112.120504}.

The CV physical noises arising from imperfect Gaussian operations, finite-energy GKP encoding, and environmental decoherence drives the bosonic state outside the logical subspace~\cite{BRADY2024100496}. Repeated GKP error correction (GKP--EC) projects the state back onto the codespace while correcting small displacement errors and digitizing continuous-variable noise into discrete logical Pauli errors~\cite{PhysRevA.102.062411, PRXQuantum.2.020101}, which can subsequently be corrected using concatenated qubit QEC codes~\cite{PhysRevA.101.012316, PRXQuantum.3.010315}.

Recently, a fault-tolerance threshold has been established against a broad class of experimentally relevant Markovian noises, including Gaussian random displacements, finite-squeezing errors, finite-resolution homodyne measurements, and optical loss~\cite{Matsuura2026}.
The digitization of the physical noises depends crucially on the implementation of GKP--EC, which can be realized either through quadrature syndrome measurements followed by corrective displacements (Steane-type EC) or, equivalently, via GKP teleportation (Knill-type EC)~\cite{marqversen2025,PRXQuantum.3.010315,PRXQuantum.2.020101}. In both approaches, the ancillary GKP states are typically assumed to be free from environmental decoherence, although they can possess finite squeezing. The impact of decohering ancillae on the GKP--EC, however, remains largely unexplored. 

In this Letter, we address this gap by studying the teleportation-based GKP--EC when the ancillary Bell modes, used as the entanglement resource~\cite{PhysRevA.102.062411}, are subject to an environmental decoherence due to the coupling to an external environment.
We employ arguably the most common bosonic decoherence mechanism, namely, the quantum Brownian-motion (QBM) model, in which the modes are linearly coupled through their position quadrature to a bath of harmonic oscillators~\cite{CALDEIRA1983587,PhysRevD.45.2843,PhysRevD.53.2012}. 
We consider a fully quantum description of the microscopic dynamics of the decoherence modeled by the system-bath interaction Hamiltonian. 
It captures a broad class of decoherence mechanisms relevant to bosonic platforms, including nanomechanical resonators~\cite{Groblacher2015}, trapped-ion motional modes~\cite{PhysRevA.69.052101,Myatt2000,PhysRevLett.97.130402}, optomechanical systems~\cite{Youssefi2023}, and superconducting microwave cavities~\cite{RevModPhys.73.357,Leger2019}. We show that such decoherence degrades the GKP Bell pair and consequently introduces imperfections into the GKP EC.

Further, we derive the teleportation Kraus operator, which defines a novel composite Pauli--Gaussian noise channel. To the best of our knowledge, this noise model has not been identified previously. We show that this correlated noise causes the encoded state to leak out of the GKP codespace, thereby invalidating the conventional mapping of CV displacement errors onto logical Pauli errors. Consequently, contrary to the conventional understanding, repeated application of GKP--EC leads to the accumulation of leakage outside the codespace, posing a fundamental challenge to scalable GKP quantum computation in the presence of environmentally induced ancilla decoherence. Counterintuitively, however, the accumulated leakage eventually saturates, with its asymptotic value determined by the decoherence strength.


\paragraph*{Notation and conventions.--}
We use the convention $\hbar=k_B=1$. For a single bosonic mode $\hat{a}$, 
$\hat{q}:=(\hat{a}+\hat{a}^\dagger)/\sqrt{2}$, 
$\hat{p}:=-i(\hat{a}-\hat{a}^\dagger)/\sqrt{2}$,
satisfying $[\hat{q},\hat{p}]=i$, and $\hat{D}(u+iv)=\exp[i(v\hat{q}-u\hat{p})]$ is the  displacement operator.
The operators
$\hat{X}(u):=\hat{D}(u)=e^{-iu\hat{p}}$ and
$\hat{Z}(v):=\hat{D}(iv)=e^{iv\hat{q}}$
generate translations by $u$ and $v$ in $\hat{q}$ and $\hat{p}$, respectively. We introduce the phase-space vector
$\mathbf{t}:=(u,v)\in\mathbb{R}^2$
and write
$\hat{D}(\mathbf{t})\equiv\hat{D}(u+iv)$ .

The GKP protocol uses the shifted Einstein--Podolsky--Rosen (EPR) states,
$\ket{EPR(\mathbf{m})}
:=({1}/{\sqrt{2\pi}})
\int dr\,e^{itr}
\ket{r}_{\hat{q}_1}
\ket{r+s}_{\hat{q}_2}$,
parameterized by phase-space vector $\mathbf{m}=(s,t)$, where $\ket{r}_{\hat{q}}$ is the position eigenstate. The EPR states form a complete basis of the two-mode Hilbert space, with the corresponding projector denoted by
$\hat{\Pi}_{EPR}(\mathbf{m}):=\ketbra{EPR(\mathbf{m})}$.

In the square-lattice GKP code, the logical Pauli-$\hat{\sigma}_z$ eigenstates
$\{|0_L\rangle,|1_L\rangle\}$
are represented by the ideal Dirac-comb states
$\ket{j_L}
=(2\sqrt{\pi})^{1/2}
\sum_{n\in\mathbb{Z}}
\ket{(2n+j)\sqrt{\pi}}_{\hat{q}}$,
where $j\in\{0,1\}$~\cite{PhysRevA.64.012310}.
Approximate finite-energy GKP states are conveniently modeled by the photon-damping operator,
$\ket{j_{\beta}}
=
N_j^{-1}(\beta)\,
e^{-\beta\hat{n}}
\ket{j_L}$,
where $\hat{n}$ is the number operator, $\beta$ denotes the damping strength, and $N_j(\beta)$ is the normalization factor~\cite{PhysRevA.102.032408,PhysRevA.102.062411}.

The logical Pauli operators are implemented by $\sqrt{\pi}$ displacements,
$\hat{X}_L=\hat{X}(\sqrt{\pi})$ and
$\hat{Z}_L=\hat{Z}(\sqrt{\pi})$.
Displacements by $2\sqrt{\pi}$ approximately preserve the finite-energy codespace and therefore act as the stabilizer operators.
We further denote the normalized two-mode GKP Bell state by
$|{\Phi_{\beta}^{+}}\rangle
\propto
|0_{\beta},0_{\beta}\rangle
+
|1_{\beta},1_{\beta}\rangle$,
and its corresponding density operator by
$\Phi_{\beta}^{+}$.
The operator $\hat{\Pi}_{\beta}:=|0_{\beta}\rangle\langle 0_{\beta}| + |1_{\beta}\rangle\langle 1_{\beta}|$ denotes the quasi-projector onto the finite-energy GKP logical codespace. Due to the finite energy constraint, $\hat{\Pi}_\beta^2\neq \hat{\Pi}_\beta$ for $\beta\neq 0$. 

\paragraph*{Standard teleportation-based GKP error correction.--}
During quantum computation, GKP-encoded modes accumulate CV noise and gradually leak out of the logical codespace. GKP--EC restores the encoded state to the codespace while converting residual CV noise into a small logical error~\cite{PhysRevA.102.062411,PRXQuantum.3.010315}.  Then the  teleportation over a GKP Bell state $|\Phi_{\beta}^{+}\rangle$ provides a natural realization of the GKP--EC by projecting 
back the input mode onto the logical codespace.

The shared Bell state can be prepared by interfering two finite-energy qunaught states,
$|\varnothing_{\beta}\rangle\propto e^{-\beta\hat{n}}\sum_{n\in\mathbb{Z}}
\ket{n\sqrt{2\pi}}_{\hat{q}}$,
on a balanced beam splitter in optical platforms, or equivalently by applying the controlled-phase gate
$CZ:=e^{-i\hat{q}_1\hat{q}_2}$
to two modes prepared in
$|+_{\beta}\rangle\propto|0_{\beta}\rangle+|1_{\beta}\rangle$
in matter-based oscillator platforms.
The teleportation protocol is implemented by performing an EPR measurement,
$\{\hat{\Pi}_{EPR}(\mathbf{m})\}_{\mathbf{m}\in\mathbb{R}^2}$,
jointly on the input mode and one mode of the shared Bell pair. Experimentally, this measurement is realized using a controlled-NOT-type entangling gate followed by complementary quadrature measurements. The resulting conditional evolution of the remaining mode is described by the Kraus operator~\cite{PhysRevA.102.062411,PRXQuantum.3.010315}
\begin{equation}\label{eq1}
    \hat{\mathcal{T}}_{\beta}(\mathbf{m})
    =
    \hat{\Pi}_{\beta}\hat{D}(-\mathbf{m}).
\end{equation}
The operator $\hat{\mathcal{T}}_{\beta}(\mathbf{m})$ first applies the displacement $\hat{D}(-\mathbf{m})$ to the input state and subsequently projects it onto the finite-energy GKP codespace. The corresponding unnormalized conditional output state is
$\bar{\rho}(\mathbf{m})
=
\hat{\mathcal{T}}_{\beta}(\mathbf{m})
\rho_{\mathrm{in}}
\hat{\mathcal{T}}_{\beta}^{\dagger}(\mathbf{m}),$
while the normalized state is
$\tilde{\rho}(\mathbf{m})
=
{\bar{\rho}(\mathbf{m})}/
{\tr[\bar{\rho}(\mathbf{m})]},$
where $\tr[\bar{\rho}(\mathbf{m})]$ gives the probability density of obtaining the EPR measurement outcome $\mathbf{m}$.

\paragraph*{Decoherence model.--}
Interactions with bosonic environments are ubiquitous in motional and resonator-based bosonic qubits~\cite{BRADY2024100496,Hanggi2005-cc,schlosshauer2007decoherence}. Although environmental coupling is generally less prominent in optical platforms, it can arise through matter impurities and matter-mediated processes~\cite{Melati14,Dutt2024,PhysRevA.105.022436,PhysRevX.13.031001}.
We model the system--bath interaction using the multi-mode QBM model, in which the system modes, here represented by the GKP Bell modes used as the teleportation resource,  couple linearly to a bath of harmonic oscillators~\cite{CALDEIRA1983587,PhysRevD.45.2843,PhysRevD.53.2012}. 
Following the standard quantum communication paradigm, we neglect the self-Hamiltonians of the communication modes, assuming their effects are already incorporated into the mode dynamics. 
This isolates the system--bath interaction, providing a tractable framework to assess how environmental decoherence of the GKP Bell pair affects the GKP--EC.

The system--bath interaction is modeled by the Hamiltonian
$\hat H_{\mathrm{int}}=(\hat{q}_A+\hat{q}_B)\otimes\sum_k g_k\hat{q}_{E_k}$,
where $\hat{q}_A$ and $\hat{q}_B$ are the quadratures of the two Bell modes, $\hat{q}_{E_k}$ is the position of the $k$th bath mode, and $g_k$ is the coupling strength. We consider here both modes coupling to the same bath. The bath self-Hamiltonian is given by $\hat{H}_E=\sum_k\omega_k(\hat{b}_k^\dagger\hat{b}_k+1/2)$, where $\omega_k$ is the angular frequency of the $k$th bath mode and $\hat{b}_k^\dagger$ ($\hat{b}_k$) is its creation (annihilation) operator. 
We take the environment initially to be in thermal equilibrium at temperature $T$, with $\rho_E=Z_E^{-1}e^{-\hat{H}_E/T}$, where $Z_E=\tr\left[e^{-\hat{H}_E/ T}\right]$ is the partition function. The system--bath coupling is characterized by the spectral density
$J(\omega)
=
\sum_k \frac{g_k^2}{2\omega_k}
\delta(\omega-\omega_k)$.  This microscopic description treats the environment as a quantum bath of harmonic modes, thereby capturing decoherence arising from the quantum interactions rather than by a phenomenological classical noise process. 

\paragraph{Kraus operator for the noisy teleportation.--} 
The model considered here admits an exact solution in the interaction picture, in which the matrix elements $\langle x,y|\Phi^+_\beta|x',y'\rangle$ acquire both a time-dependent, bath-induced
phase $e^{i\phi(t)\left[(x+y)^2-(x'+y')^2\right]}$ and the decoherence factor $\Gamma_t(x,y,x',y')$~\cite{SM}. The quadratic phase corresponds to a deterministic Gaussian unitary evolution, equivalent to a shear map, and can therefore be corrected or compensated for through the mode dynamics. We thus omit this deterministic contribution and focus on the decoherence effects, governed by the decoherence factor (see~\cite{SM}) 
\begin{equation}
\begin{aligned}
    \Gamma_t(x,y,x',y')
& = 
\exp\!\left[
-\tau(t)\left\{(x'-x)+(y'-y)\right\}^2
\right],\\
\end{aligned}
\end{equation}
which reflects the irreversible effects of the environment on the resource. Above (see~\cite{SM} for the derivation)
\begin{equation} \label{tau}
    \begin{aligned}
        \tau(t)
    & =
    \frac{1}{\pi}\int_0^\infty d\omega\,
    \frac{J(\omega)}{\omega^2}
    \left[1-\cos(\omega t)\right]
    \coth\left(
        \frac{\omega}{2 T}
    \right), \\
    \end{aligned}
\end{equation}
is the interaction-time related factor, determined also by the spectral density $J(\omega)$ and the bath temperature.

\begin{figure*}[t]
    \centering
\begin{minipage}{0.31\textwidth}
    \centering
    \includegraphics[width=0.95\linewidth]{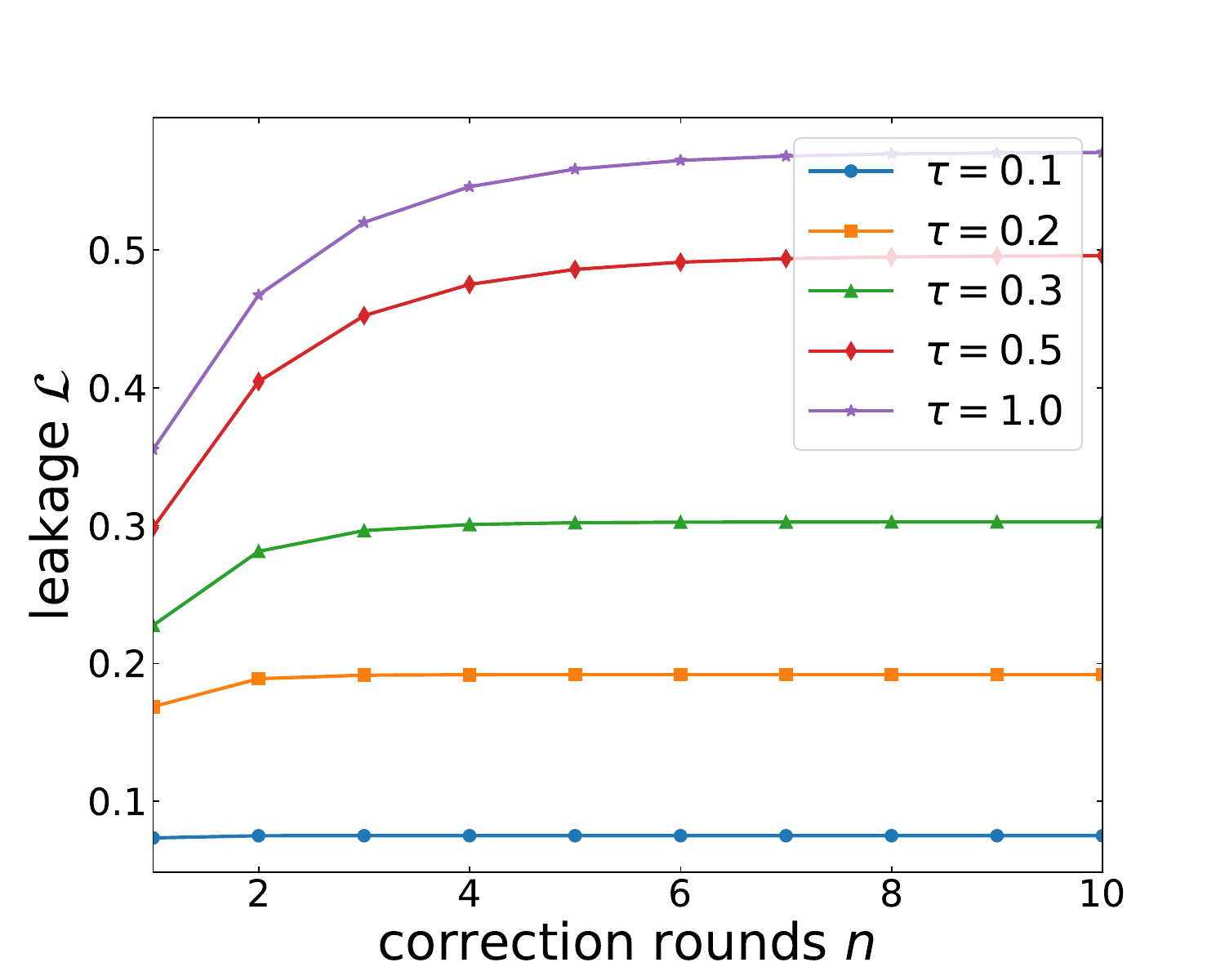}\\[-1pt]
    (a)
\end{minipage}
\hspace{0.01\textwidth}
\begin{minipage}{0.31\textwidth}
    \centering
    \includegraphics[width=0.95\linewidth]{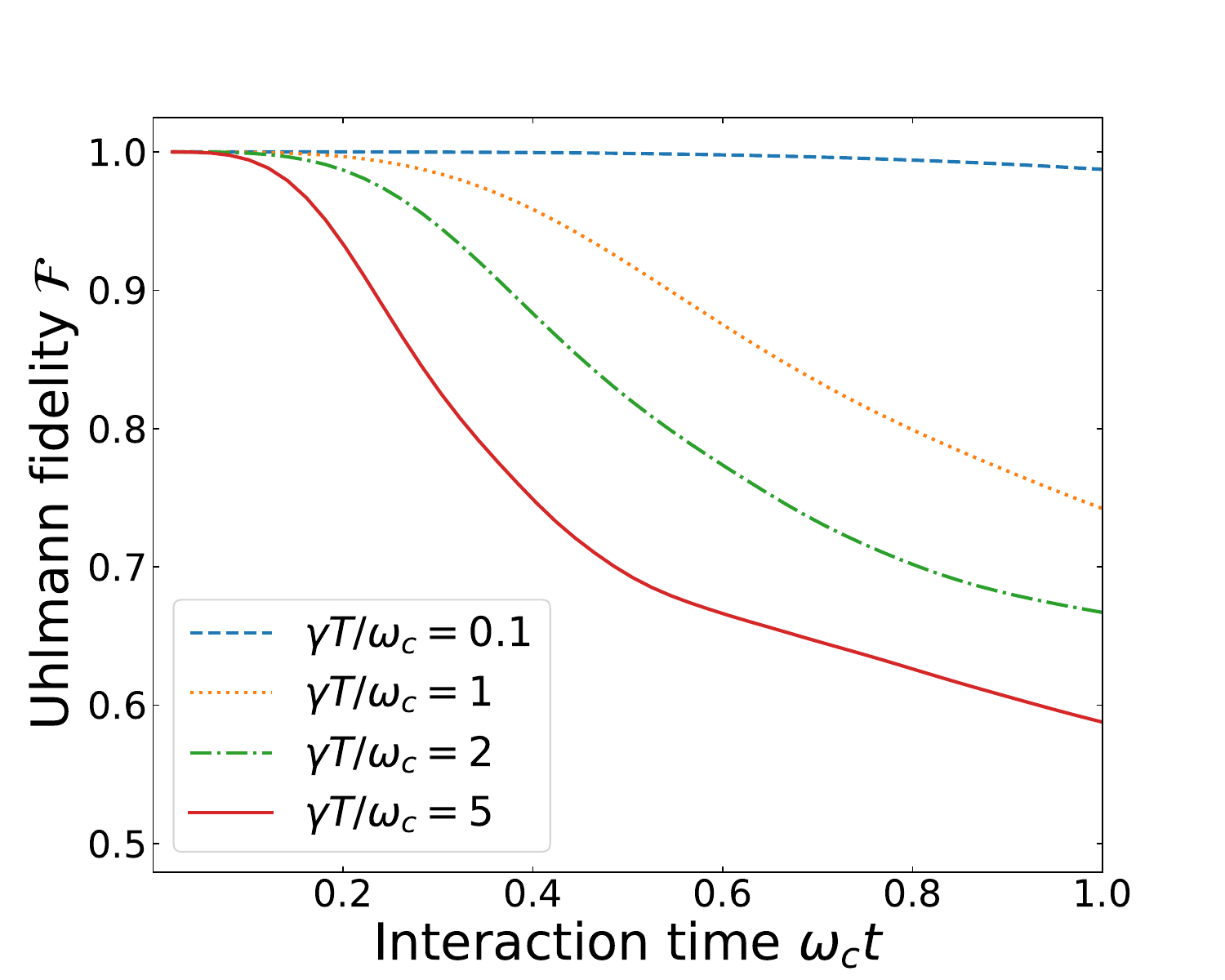}\\[-1pt]
    (b)
\end{minipage}
\hspace{0.01\textwidth}
\begin{minipage}{0.31\textwidth}
    \centering
    \includegraphics[width=0.95\linewidth]{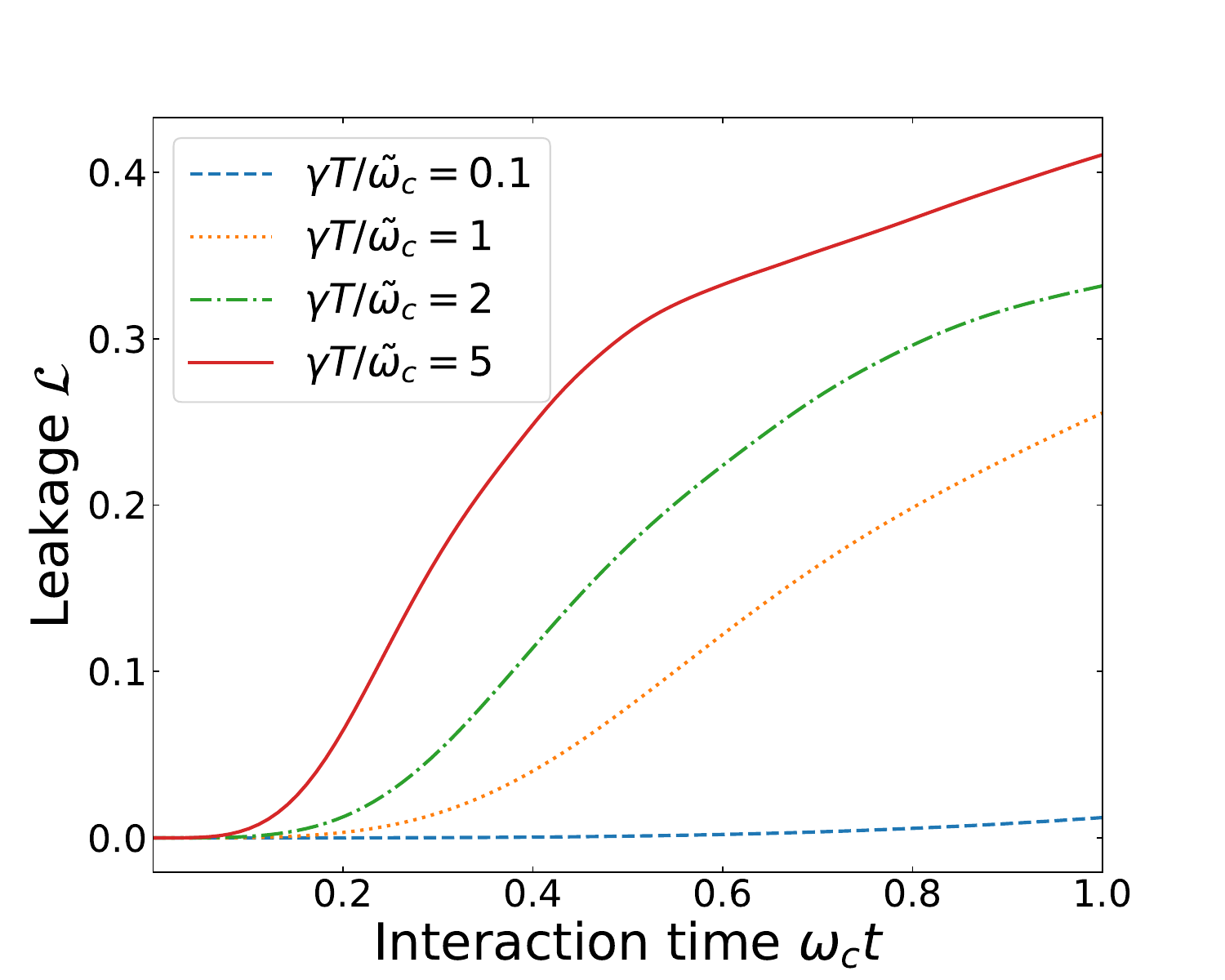}\\[-1pt]
    (c)
\end{minipage}

\includegraphics[width=0.95\linewidth]{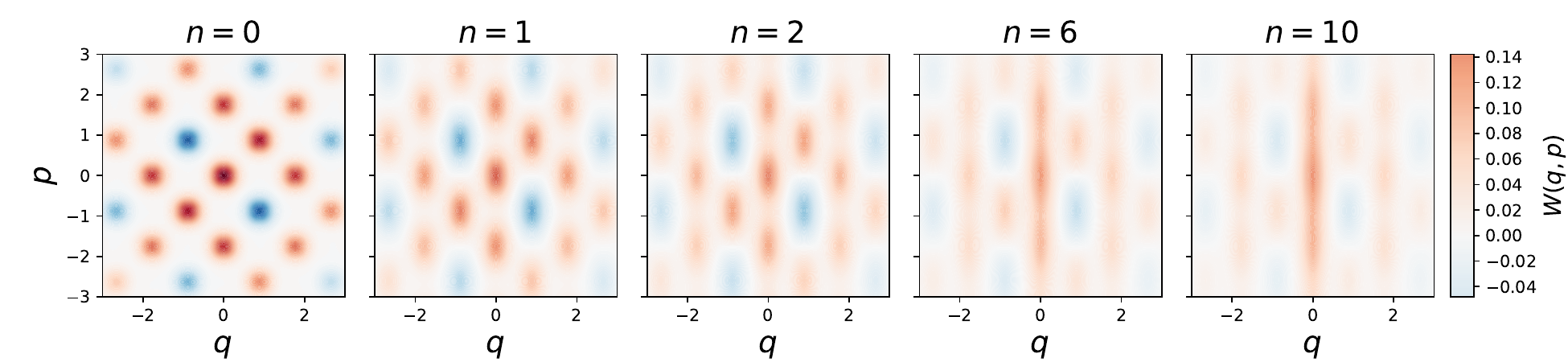}\\[-1pt]
(d)
    \caption{%
    (a) Leakage out of the GKP subspace as a function of successive GKP--EC rounds, averaged over the logical Pauli basis states for varying $\tau$. Homodyne outcomes are fixed to $(0,0)$ in each round, so that no unitary correction is applied. (b) Uhlmann fidelity of teleportation as a function of the dimensionless time $\omega_c t$ for the input state $\ket{0_\beta}$, with homodyne outcomes $(0,0)$, for varying values of $\gamma T/\omega_c$ and a Lorentz--Drude cutoff, Eq.~\eqref{tau_LD}. The results for the exponential cutoff, Eq.~\eqref{tau_exp}, are qualitatively similar. (c) Leakage for the same parameters as in (b). (d) Wigner functions of the $+1$ eigenstate of the GKP Pauli operator $\hat{\sigma}_y$, with $\beta=0.1$ and $\tau=0.5$, after $n=0,1,2,6,10$ successive teleportation rounds. 
    }
    \label{fig:results}
\end{figure*}

The Kraus operator corresponding to the teleportation through a decohered GKP Bell state, without applying the Pauli correction, is given by~\cite{SM}
\begin{equation}\label{eq4}\hat{\mathcal{E}}_{\mathbf{m},\tau(t)}(\cdot)
    =
    \int dr\,\mathcal{G}_{0,\tau(t)}(r)
\hat{\mathcal{K}}_{\mathbf{m}}(r)(\cdot)\hat{\mathcal{K}}^\dagger_{\mathbf{m}}(r),
\end{equation}
where $\mathcal{G}_{0,\tau(t)}(r) = (1/\tau\sqrt{8\pi})e^{-{r^2}/{4\tau(t)^2}}$ is the zero-mean Gaussian distribution with the standard deviation $\sqrt{2}\tau(t)$, characterizing the decoherence-induced random displacement, and
\begin{equation} \label{Eq:5}
    \hat{\mathcal{K}}_{\mathbf{m}}(r)
    =
    \hat{D}(-ir)\,
    \hat{\mathcal{T}}_\beta(\mathbf{m})\,
    \hat{D}(-ir).
\end{equation}
is a randomized Kraus operator acting on the input state. Here, $\mathbf{m}$ denotes the EPR measurement outcome, while $\hat{\mathcal{T}}_\beta(\mathbf{m})$ is given by Eq.\eqref{eq1}. The input state is randomized twice by $\hat{\mathcal{E}}_{\mathbf{m},\tau}$: before GKP projection, producing an $r$-dependent Pauli error, and after projection, introducing continuous displacement noise. The same random variable $r$ enters coherently on both sides of the ideal finite-energy teleportation Kraus operator, yielding a correlated Pauli–Gaussian noise channel. The Gaussian displacement noise broadens the GKP peaks and can therefore be interpreted as a degradation of the effective GKP squeezing, although it is not equivalent to a bosonic damping channel~\cite{PhysRevA.102.032408,Hastrup2023}. 

\paragraph{Analysis of the noisy protocol.--} 
The noisy teleportation channel, Eq.~\eqref{eq4}, corrupts the GKP--EC: Instead of correcting the displacement errors, it introduces additional displacement noise, resulting in leakage from the logical subspace. We quantify this leakage as $\mathcal{L}(\rho):=1-\tr[\rho\hat{P}_\beta]$, where $\hat{P}_\beta:=\ketbra{\tilde{0}_\beta}+\ketbra{\tilde{1}_\beta}$ is the projector constructed using the Gram--Schmidt orthogonalization of the finite-energy GKP basis, with $\ket{\tilde{0}_\beta}=\ket{0_\beta}$ and $\ket{\tilde{1}_\beta}\propto\ket{1_\beta}-\braket{0_\beta}{1_\beta}\ket{0_\beta}$. 
Because the noisy output is not confined to the finite-energy logical subspace, the usual two-dimensional logical-state fidelity does not capture the full error~\cite{Hastrup2023}. We therefore use leakage as a complementary measure of the CV component of the noise. Residual leakage from one GKP--EC step is carried over to the next, driving the mode progressively further from the logical subspace. This leads to an accumulation of leakage over successive EC rounds.

FIG.~\ref{fig:results}(a) shows the leakage accumulated over $n$ successive GKP teleportation rounds, $\mathcal{L}_n$, as a function of 
$n$. We compute $\mathcal{L}_n$ numerically using Eq.~\eqref{eq4}, averaging over the logical Pauli basis states. FIG.~\ref{fig:results}(d) plots the Wigner function for the logical Pauli basis $(\ket{0_\beta}+i\ket{1_\beta})/\sqrt{2}$ for different number of successive GKP--EC rounds. The logical information is quickly destroyed by the noisy teleportation. We use the QuTiP Python framework~\cite{Johansson2012QuTiP} for the numerics and set the Fock cutoff to $50$ for $\beta=0.1$, corresponding to approximately $10\;\mathrm{dB}$ of mode squeezing. The leakage increases with the number of rounds but, surprisingly, approaches a plateau characterized by the asymptotic value $\mathcal{L}_{\infty}$. We phenomenologically model this coarse-grained evolution, consistent with the simulated asymptotic saturation of FIG.~\ref{fig:results}(a), 
by
$\mathcal{L}_{n+1}\simeq\mathcal{L}_{n}+\lambda\left(\mathcal{L}_{\infty}-\mathcal{L}_{n}\right),$
where $0\leq\lambda<1$ characterizes the leakage accumulation per round, and $\mathcal{L}_{\infty}$ is the asymptotic leakage determined by $\beta$ and $\tau$.
For small $\tau$, however, $\mathcal{L}_{\infty}$ is negligible, rendering leakage accumulation insignificant. This can be understood intuitively from $\hat{\mathcal{T}}_\beta(\mathbf{m})$: The quasi-projector $\hat{\Pi}_\beta$ corrects the small random displacement introduced by the rightmost $\hat{D}(-ir)$ in Eq.~\eqref{Eq:5}, while the leftmost $\hat{D}(-ir)$ induces only a small randomization narrowly centered around the identity. This advantage diminishes as $\tau$ increases.

Due to the leakage, the logical fidelity is no longer an appropriate metric for benchmarking the teleportation performance. We therefore use the Uhlmann fidelity, denoted by $\mathcal{F}$, instead~\cite{Jozsa01121994,UHLMANN1976273}. For a pure input state $|\psi\rangle$, the Uhlmann fidelity of the teleportation channel is given by $\mathcal{F}=\langle\psi|\mathcal{E}_{\mathbf{m},\tau(t)}(|\psi\rangle\langle\psi|)|\psi\rangle$.

\paragraph{Realistic environments.--} To connect to realistic environments, we consider an Ohmic spectral density $J(\omega)$ with two commonly used spectral cutoffs: (i) Lorentz--Drude and (ii) exponential~\cite{schlosshauer2007decoherence, BreuerPetruccione2002}. We then use realistic parameters reported in the experimental literature for quantum oscillators proposed as potential platforms for GKP qubits to estimate $\tau$. For the Lorentz--Drude cutoff, the Ohmic spectral density is
\begin{equation}
J(\omega)
=
\frac{
2\gamma\,\omega\,\omega_c^2
}{
\omega^2+\omega_c^2
},
\end{equation}
where, $\gamma$ is the effective system--bath coupling strength, while $\omega_c$ is the 
cutoff frequency. For motional oscillators operated at cryogenic temperatures, the thermal energy scale can still be much larger than the oscillator energy scale. For example, at $T=4\;\mathrm{K}$, $k_BT/\hbar\simeq83.35\;\mathrm{GHz}$, whereas the oscillator frequencies are typically in the range of $100\;\mathrm{kHz}$--$10\;\mathrm{MHz}$. Thus, even at cryogenic temperatures, the relevant systems lie in the high-temperature regime, $T\gg\omega$, where we approximate $\coth(\omega/2T)\approx2T/\omega$ in Eq.~\eqref{tau}, which gives~\cite{SM}:
\begin{equation} \label{tau_LD}
\tau_{LD}
=
\frac{2\gamma T}{\omega_c}
\left[
\omega_c t - 1 + e^{-\omega_c t}
\right].
\end{equation}
The combination $\bar n=\gamma T/\omega_c$ sets the phenomenological motional heating rate and is related to the mean thermal occupation, 
$n_{\rm th}=1/(e^{\omega_c/T}-1)\simeq T/\omega_c$. Since the heating rate can be directly inferred from experiments, $\tau$ can be estimated using experimentally accessible parameters rather than relying on microscopic modeling of the system--bath coupling. Thus, our approach reduces the characterization of decoherence to the empirical parameters $\gamma T$ and $\omega_c$, providing a common framework for diverse GKP platforms subject to Lorentz--Drude thermalization. The resulting Uhlmann fidelity and the leakage are presented in FIG.~\ref{fig:results}(b) and (c), respectively, showing that the teleportation fidelity decreases/the leakage increases rapidly  with the increasing heating rate $\gamma T/\omega_c$, even at short times $\omega_c t$. 

For the exponential cutoff, $J(\omega)=\pi\gamma\omega e^{-\omega/\omega_c}$, in the same high-temperature regime, Eq.~\eqref{tau} yields~\cite{SM}
\begin{equation}\label{tau_exp}
\tau_{exp}
=
\frac{2\gamma T}{\omega_c} \left[(\omega_ct)\tan^{-1}(\omega_c t)-
\frac{\ln\left(1+\omega_c^2t^2\right)}{2}\right].
\end{equation}
For short times $\omega_c t\;\ll 1$, $\tau_{LD}$ and $\tau_{exp}$ behave similarly, scaling quadratically in time, while for long times the resulting decoherence is algebraic rather than exponential.

\paragraph{Estimation for realistic GKP platforms.--} Here, we estimate $\tau$ for realistic trapped-ion-based platforms reported in recent years.
The motional heating rate $\bar{n}$ is measured experimentally in units of quanta per second. The analysis of $\tau$ can therefore be reduced to a function of $\bar{n}$ and the dimensionless time parameter $\omega_c t$.
For illustrative estimates we assume $\omega_c=10\Omega$, a commonly employed separation of bath and system frequency scales in QBM modeling~\cite{Paavola2009,Piilo2007}. A recently reported cryogenic $^{40}\mathrm{Ca}^{+}$ ion trap exhibits dielectric-induced heating at a rate of $\bar{n}\simeq 30\,\mathrm{s}^{-1}$ for a motional frequency of $1.5$ MHz~\cite{wj8m-1nlq}. A $^{9}\mathrm{Be}^{+}$ Penning microtrap, on the other hand, demonstrates an ultralow heating rate of $\bar{n}\simeq 0.09\,\mathrm{s}^{-1}$ at a motional frequency of $2.5$ MHz~\cite{Jain2024}. In contrast, room-temperature ion traps can exhibit substantially higher heating rates, of order $10^3\,\mathrm{s}^{-1}$ for similar motional frequencies, which may render them unsuitable for GKP encoding because of the associated short coherence times~\cite{Chung_2025}. A particularly relevant experiment, directly related to trapped-ion GKP Bell-state preparation and universal gate implementation, was recently reported using a single ${}^{171}\mathrm{Yb}^{+}$ ion confined in a room-temperature Paul trap~\cite{Matsos2025}. The authors report motional frequencies of order $\sim 10\,\mathrm{MHz}$, a heating rate of $\sim 0.2\,\mathrm{s}^{-1}$, and gate durations of $\sim 200\mathbin{-}300\,\mu\mathrm{s}$. These recent experimental developments suggest an achievable low heating rate in the range $\sim 0.01\mathbin{-}1\,\mathrm{s}^{-1}$, $\omega_c\simeq 10\mathbin{-}100\,\mathrm{MHz}$, and gate durations of order $\sim 100\mathbin{-}1000\,\mu\mathrm{s}$. These parameters translate to $\tau\simeq 10\mathbin{-}10^5\gg 1$, equivalently $\mathcal{L}\simeq 0.39\mathbin{-}0.60$ or $\mathcal{F}\simeq 0.55\mathbin{-}0.36$ for $\beta=0.1$ in the first round of teleportation, resulting in a rapid accumulation of leakage out of the logical subspace when successive rounds of GKP--EC are performed. This substantial leakage accumulation therefore poses a serious challenge to the practical realization of FTQC with trapped-ion architectures based on teleportation-based GKP--EC.


\paragraph{Summary and outlook.--} In this Letter, we have addressed the crucial question of how teleportation-based GKP--EC is affected when the GKP Bell-state ancilla has undergone pure dephasing. We have shown that the corresponding teleportation Kraus operators introduce additional CV errors into the logical modes, rather than correcting or digitizing the residual errors propagated from preceding GKP--EC steps or logical operations. The resulting noise exhibits a previously unreported characteristic: correlated Pauli-Gaussian noise that drives the input mode outside the logical subspace. Under repeated teleportation, this leakage accumulates and can potentially lead to the complete destruction of the encoded logical information. To model the decoherence, we considered the pure-dephasing limit of QBM and employed an Ohmic spectral density with Lorentz-Drude and exponential cutoffs. These results indicate that environmental decoherence of the teleportation resource states can constitute an important error mechanism in motional-oscillator implementations and should be included in fault-tolerance analyses.

Although our analysis uses the QBM model as an example, it provides a framework for investigating other sources of decoherence-induced noise that are ubiquitous in bosonic systems, such as dephasing arising from fast fluctuations of the trapping potential or thermal fluctuations~\cite{Sivak2023,schlosshauer2007decoherence}. Extending the present analysis to these noise mechanisms and deriving the corresponding teleportation Kraus operators would be interesting. 
For concreteness, we have assumed here an Ohmic spectral density 
but more general spectral densities, including sub- and super-Ohmic, inverse-power-law frequency dependences, and boson-spin interactions could be straightforwardly incorporated as well to capture different bath characteristics and environmental couplings~\cite{spectraldensities,Wolfowicz2021,RevModPhys.86.361}.

Another avenue aligned with this work is to investigate the full fault tolerance within the recently established stabilizer--subsystem framework for GKP codes~\cite{PRXQuantum.5.010331}, and to determine the fault-tolerance thresholds for different bosonic platforms while accounting for platform-specific decoherence mechanisms and the resulting noise in the Bell-pair ancillae. Systematically comparing these thresholds across architectures could help identify platforms capable of supporting scalable GKP-based FTQC.


Last but not least, our results call for a design of a passive or active protection/correction mechanism against the pure-dephasing noise, e.g. along the lines of the purifying teleportation effect~\cite{Roszak2023purifying}.
 
We acknowledge the support of National Science Centre (NCN) through the QuantEra project QuCABOoSE 2023/05/Y/ST2/00139.

\bibliographystyle{apsrev4-1}
\bibliography{ref}

\pagebreak





\clearpage
\newpage
\onecolumngrid
\begin{center}
	\textbf{\Large Supplemental Materials}
\end{center}

\setcounter{equation}{0}
\setcounter{figure}{0}
\setcounter{table}{0}
\setcounter{page}{1}
\makeatletter
\renewcommand{\theequation}{S\arabic{equation}}
\renewcommand{\thefigure}{S\arabic{figure}}
\renewcommand{\bibnumfmt}[1]{[S#1]}
\renewcommand{\citenumfont}[1]{S#1}

\section{I. Derivation of the decoherence factor}

\subsection{A. Pure-dephasing limit of quantum Brownian motion}

We consider a bosonic system mode whose self-Hamiltonian is neglected,
$H_S=0$, while retaining the dynamics of the bosonic environment. The
Hamiltonian is therefore
\begin{equation}
    H = H_E + H_{\mathrm{int}},
\end{equation}
where
\begin{equation}
\hat{H}_E=\sum_k\omega_k(\hat{b}_k^\dagger\hat{b}_k+1/2),
\end{equation}
or equivalently 
\begin{equation}
    \hat{H}_E =
    \frac{1}{2}\sum_k
    \left(
        {\hat p_k^2}
        + {\omega_k^2\hat q_k^2}
    \right),
\end{equation}
and the system--bath interaction is taken to be linear in the system
position,
\begin{equation}
    H_{\mathrm{int}}
    =
    (\hat{q}_A+\hat{q}_B)\otimes\sum_k g_k\hat{q}_{E_k}.
\end{equation}
This corresponds to the $H_S=0$ (pure-dephasing) limit of the
quantum Brownian-motion model. Since the system Hamiltonian vanishes, $\hat q$ is invariant in the
interaction picture,
\begin{equation}
    \hat q_I(t)=\hat q.
\end{equation}
The bath quadratures evolve according to
\begin{equation}
    \hat q_k(t)
    =
    {\frac{1}{\sqrt{2}}}
    \left(
        \hat b_k e^{-i\omega_k t}
        +
        \hat b_k^\dagger e^{i\omega_k t}
    \right),
\end{equation}
and hence
\begin{equation} \label{H_int_pic}
    H_I(t)
    =
    \hat (\hat q_A + \hat q_B)\otimes
    \sum_k \frac{g_k}{\sqrt{2}}
    \left(
        \hat b_k e^{-i\omega_k t}
        +
        \hat b_k^\dagger e^{i\omega_k t}
    \right).
\end{equation}
This gives the interaction-picture evolution operator as
\begin{equation}
    U_I(t)
    =
    \mathcal{T}
    \exp\left[
        -\frac{i}{\hbar}
        \int_0^t ds\,H_I(s)
    \right],
\end{equation}
which can be further expressed as Magnus expansion as
\begin{equation}
    U_I(t) = \mathcal{T}\exp\left[\Omega_1(t) + \Omega_2(t) + \Omega_3 (t) + \cdots \right].
\end{equation}
The first two Magnus terms are
\begin{equation}\label{magnus_1}
    \Omega_1(t)
    =
    -i\int_0^t ds\,\hat{H}_{I}(s),
\end{equation}
and
\begin{equation}\label{magnus_2}
    \Omega_2(t)
    =
    -\frac{1}{2}
    \int_0^t ds_1
    \int_0^{s_1} ds_2\,
    [\hat{H}_{I}(s_1),\hat{H}_{I}(s_2)].
\end{equation}
Higher-order terms contain nested commutators.
By denoting $\hat{q}_A+\hat{q}_B\equiv \hat{Q}$, and using Eqs.~\eqref{H_int_pic} and \eqref{magnus_1}, we have
\begin{equation}
\begin{aligned}
    \Omega_1(t)
    &=
    -i\hat{Q}
    \sum_k
    \frac{g_k}{\sqrt{2}}
    \int_0^t ds
    \left(
        \hat{b}_{k}e^{-i\omega_k s}
        +
        \hat{b}_k^{\dagger}e^{i\omega_k s}
    \right)\\
    & = \hat{Q}
    \sum_k
    \frac{g_k}{\sqrt{2}\,\omega_k}
    \left[
        \left(e^{-i\omega_k t}-1\right)\hat{b}_{k}
        +
        \left(1-e^{i\omega_k t}\right)\hat{b}_k^{\dagger}
    \right], \\
    & =
    \hat{Q}
    \sum_k
    \left[
        \alpha_k(t)\hat{b}_k^{\dagger}
        -
        \alpha_k^{*}(t)\hat{b}_{k}
    \right].
\end{aligned}
\end{equation}
where $\alpha_k(t)
    =
    \frac{g_k}{\sqrt{2}\,\omega_k}
    \left(1-e^{i\omega_k t}\right)$.
For second order term of Magnus expansion, we first calculate $[\hat{H}^{I}(s_1),\hat{H}^{I}(s_2)]$.
Since $\hat{Q}$ acts only on the system,
\begin{equation}
    [\hat{H}^{I}(s_1),\hat{H}^{I}(s_2)]
    =
    \hat{Q}^{2}
    [\hat{B}(s_1),\hat{B}(s_2)],
\end{equation}
where
\begin{equation}
    \hat{B}(s)
    =
    \sum_k
    \frac{g_k}{\sqrt{2}}
    \left(
        \hat{b}_{k}e^{-i\omega_k s}
        +
        \hat{b}_k^{\dagger}e^{i\omega_k s}
    \right).
\end{equation}
Using $[\hat{b}^{k},\hat{b}^{l\dagger}]
    =
    \delta_{kl}
    \; \mathrm{and} \;
    [\hat{b}^{k},\hat{b}^{l}]
    =
    0,$
we get
\begin{equation}
\begin{aligned}
    [\hat{B}(s_1),\hat{B}(s_2)]
    & =
    \frac{1}{2}
    \sum_k g_k^2
    \left[
        e^{-i\omega_k(s_1-s_2)}
        -
        e^{i\omega_k(s_1-s_2)}
    \right],\\
    & = -i
    \sum_k g_k^2
    \sin\left[\omega_k(s_1-s_2)\right].
\end{aligned}
\end{equation}
Therefore,
\begin{equation}
    [\hat{H}_{I}(s_1),\hat{H}_{I}(s_2)]
    =
    -i\hat{Q}^{2}
    \sum_k g_k^2
    \sin\left[\omega_k(s_1-s_2)\right].
\end{equation}
Consequently, using Eq.~\eqref{magnus_2}
\begin{equation}
\begin{aligned}
    \Omega_2(t) &
    =
    \frac{i\hat{Q}^{2}}{2}
    \sum_k g_k^2
    \int_0^t ds_1
    \int_0^{s_1} ds_2\,
    \sin\left[\omega_k(s_1-s_2)\right],\\
    & = i\hat{Q}^{2}
    \sum_k
    \frac{g_k^2}{2\omega_k^2}
    \left[
        \omega_k t-\sin(\omega_k t)
    \right], \\
    & = i\phi(t)\hat{Q}^{2}.
\end{aligned}
\end{equation}
where 
\begin{equation}
    \phi(t)
    =
    \sum_k
    \frac{g_k^2}{2\omega_k^2}
    \left[
        \omega_k t-\sin(\omega_k t)
    \right].
\end{equation}

The higher order terms of the Magnus expansion are nested commutators and because 
$[\hat{Q},\hat{Q}^{2}]=0$, while the commutator of the
interaction Hamiltonians has no remaining bath operators, we have
\begin{equation}
    [\hat{H}_{I}(s),
    [\hat{H}_{I}(s_1),\hat{H}_{I}(s_2)]]
    =0.
\end{equation}

Therefore,
\begin{equation}
    \Omega_n(t)=0,
    \qquad n\geq 3.
\end{equation}
Since $\Omega_1(t)$ and $\Omega_2(t)$ commute, $[\Omega_1(t),\Omega_2(t)]=0$,
we can write
\begin{equation}
    \hat{U}_{I}(t)
    =
    e^{i\phi(t)\hat{Q}^{2}}
    \prod_k
    \exp\left[
        \hat{Q}
        \left(
            \alpha_k\hat{b}_k^{\dagger}
            -
            \alpha_k^{*}\hat{b}_{k}
        \right)
    \right].
\end{equation}
Equivalently, introducing the displacement operator
\begin{equation}
    \hat{D}_{k}(\xi)
    =
    \exp\left(
        \xi\hat{b}_k^{\dagger}
        -
        \xi^{*}\hat{b}_{k}
    \right),
\end{equation}
the unitary $\hat{U}_{I}(t)$ can be expressed as a product of conditional
displacement operators acting on the bath,
\begin{equation}
\boxed{\hat{U}_{I}(t)
    =
    e^{i\phi(t)(\hat{q}_A + \hat{q}_B)^{2}}
    \prod_k
    \hat{D}_{k}\!\left(
        (\hat{q}_A + \hat{q}_B)\alpha_k(t)
    \right).}
\end{equation}
Thus, a two-mode system position eigenstate $|x,y\rangle$ becomes correlated with
an environment state displaced by an amount proportional to $x+y$,
\begin{equation}
    U_I(t)|x,y\rangle|\psi_B\rangle
    =
    e^{i\phi(t)(x+y)^2}
    |x,y\rangle
    \prod_k
    D_k\left((x+y)\alpha_k(t)\right)
    |\psi_B\rangle.
\end{equation}
The phase factor in $\hat{U}_I(t)$ is a deterministic Gaussian operator and can be absorbed into the self Hamiltonian of the systems or can be corrected. We limit our analysis to the part that introduces unrecoverable decoherence:
\begin{equation}
\boxed{\hat{U}_{I}(t)
    =
    \prod_k
    \hat{D}_{k}\!\left(
        (\hat{q}_A + \hat{q}_B)\alpha_k(t)
    \right).}
\end{equation}

\subsection{B. Thermal-bath decoherence factor}

We assume that initially the system and bath are uncorrelated,
\begin{equation}
    \rho(0)
    =
    \rho_S(0)\otimes\rho_E,
\end{equation}
where the bath is in a thermal state,
\begin{equation}
    \rho_E
    =
    \frac{e^{-H_E/T}}
    {\mathrm{Tr}_E(e^{-H_E/T})}.
\end{equation}
Tracing out the bath gives
\begin{equation}
    \langle x,y|\rho_S(t)|x^\prime, y^\prime\rangle
    =
    \langle x,y|\rho_S(0)|x',y'\rangle
    \Gamma(x,y,x',y';t),
\end{equation}
where the decoherence factor is
\begin{equation}
    \Gamma(x,y,x',y';t)
    =
    \mathrm{Tr}_E
    \left[
        \prod_k
        D_k\!\left((x+y)\alpha_k(t)\right)
        \rho_E
        \prod_k
        D_k^\dagger\!\left((x'+y')\alpha_k(t)\right)
    \right].
\end{equation}

For a thermal Gaussian bath, this trace can be evaluated exactly using
the thermal characteristic function of the displacement operator,
\begin{equation}
    \mathrm{Tr}
    \left[
        D(\alpha)\rho_{E}
    \right]
    =
    \exp\left[
        -\frac{1}{2}
        |\alpha|^2
        \coth\left(
            \frac{\omega}{2T}
        \right)
    \right].
\end{equation}
Consequently,
\begin{equation} \label{decoh_factor}
    \boxed{
    \Gamma(x,y,x',y';t)
    =
    \exp\left[
        -\tau(t)((x-x')+(y-y'))^2
    \right],
    }
\end{equation}
where 
\begin{equation}
   \tau(t) = \sum_k
        \frac{g_k^2}{2\omega_k^2}
        \left(
            1-\cos\omega_k t
        \right)
        \coth\left(
            \frac{\omega_k}{2 T}
        \right)
\end{equation}
Introducing the spectral density
\begin{equation}
    J(\omega)
    =
    \sum_k
    \frac{g_k^2}{2\omega_k}
    \delta(\omega-\omega_k),
\end{equation}
and using
\begin{equation}
    \sum_k
    \frac{g_k^2}{2\omega_k^2}
    f(\omega_k)
    =
    \frac{1}{\pi}
    \int_0^\infty d\omega\,
    \frac{J(\omega)}{\omega}
    f(\omega).
\end{equation}
we get
\begin{equation} \label{tau_SM}
   \boxed{\tau(t) = \frac{1}{\pi}\int_0^\infty d\omega\,
    \frac{J(\omega)}{\omega^2}\left[
        1-\cos(\omega t)
    \right]
    \coth\left(
        \frac{\omega}{2 T}
    \right). }  
\end{equation}

\subsection{C. Ohmic spectral density with Lorentz-Drude cutoff}
We consider the Lorentz--Drude spectral density

$$
J(\omega)
=
\frac{2\gamma\,\omega\,\omega_c^2}
{\omega^2+\omega_c^2}.
$$
Substitution into Eq.~\eqref{tau_SM} gives
\begin{equation}
    \tau_{LD}(t)
=
\frac{2\gamma\omega_c^2}{\pi}
\int_0^\infty d\omega\,
\frac{1-\cos(\omega t)}
{\omega(\omega^2+\omega_c^2)}
\coth\left(\frac{\omega}{2T}\right).
\end{equation}
For the high-temperature regime,
$T\gg\omega,$
we use
\begin{equation}
    \coth\left(\frac{\omega}{2T}\right)
\simeq
\frac{2T}{\omega},
\end{equation}
which yields
\begin{equation}
    \tau_{LD}(t)
\simeq
\frac{4\gamma T\omega_c^2}{\pi}
\int_0^\infty d\omega\,
\frac{1-\cos(\omega t)}
{\omega^2(\omega^2+\omega_c^2)}.
\end{equation}
Using
\begin{equation}
    \int_0^\infty d\omega\,
\frac{1-\cos(\omega t)}
{\omega^2(\omega^2+\omega_c^2)}
=
\frac{\pi}{2\omega_c^2}
\left[
t-\frac{1-e^{-\omega_c t}}{\omega_c}
\right],
\end{equation}
we obtain
\begin{equation}
\boxed{\tau_{LD}(t)
=
\frac{2\gamma T}{\omega_c}
\left[
 \omega_c t - 1 + e^{-\omega_c t}
\right].}
\end{equation}

\subsection{D. Ohmic spectral density with exponential cutoff}
The Ohmic spectral density with an exponential cutoff $J(\omega)=\pi\gamma\,\omega\,e^{-\omega/\omega_c}$ in high temperature regime $T\gg \omega_c$ gives
\begin{equation}
\tau_{exp}(t)
=
\gamma T\int_0^\infty d\omega\,
\frac{e^{-\omega/\omega_c}}{\omega^2}
\left(1-\cos\omega t\right).
\end{equation}
Differentiating with respect to $t$ gives
\begin{equation}
\frac{d\tau_{exp}(t)}{dt}
=
\gamma T\int_0^\infty d\omega\,
\frac{e^{-\omega/\omega_c}}{\omega}
\sin(\omega t).
\end{equation}
Differentiating once more,
\begin{equation}
\frac{d^2\tau_{exp}(t)}{dt^2}
=
\gamma T\int_0^\infty d\omega\,
e^{-\omega/\omega_c}
\cos(\omega t).
\end{equation}
Using
\begin{equation}
\int_0^\infty d\omega\,
e^{-a\omega}\cos(\omega t)
=
\frac{a}{a^2+t^2},
\end{equation}
with $a=1/\omega_c$, we find
\begin{equation}
\frac{d^2 \tau_{exp}(t)}{dt^2}
=
\frac{\gamma\, T\,\omega_c}{1+\omega_c^2 t^2}.
\end{equation}
Integrating once and using $\tau_{exp}'(0)=0$,
\begin{equation}
\frac{d\tau_{exp}(t)}{dt}
=
\gamma T\tan^{-1}(\omega_c t).
\end{equation}
Integrating again and using $\tau_{exp}(0)=0$,
\begin{equation}
\boxed{
\tau_{exp}(t)
=
\frac{2\gamma T}{\omega_c} \left[\omega_ct\tan^{-1}(\omega_c t)-
\frac{\ln\left(1+\omega_c^2t^2\right)}{2}\right].
}
\end{equation}

\begin{figure}
\centering\includegraphics[width=0.6\linewidth]{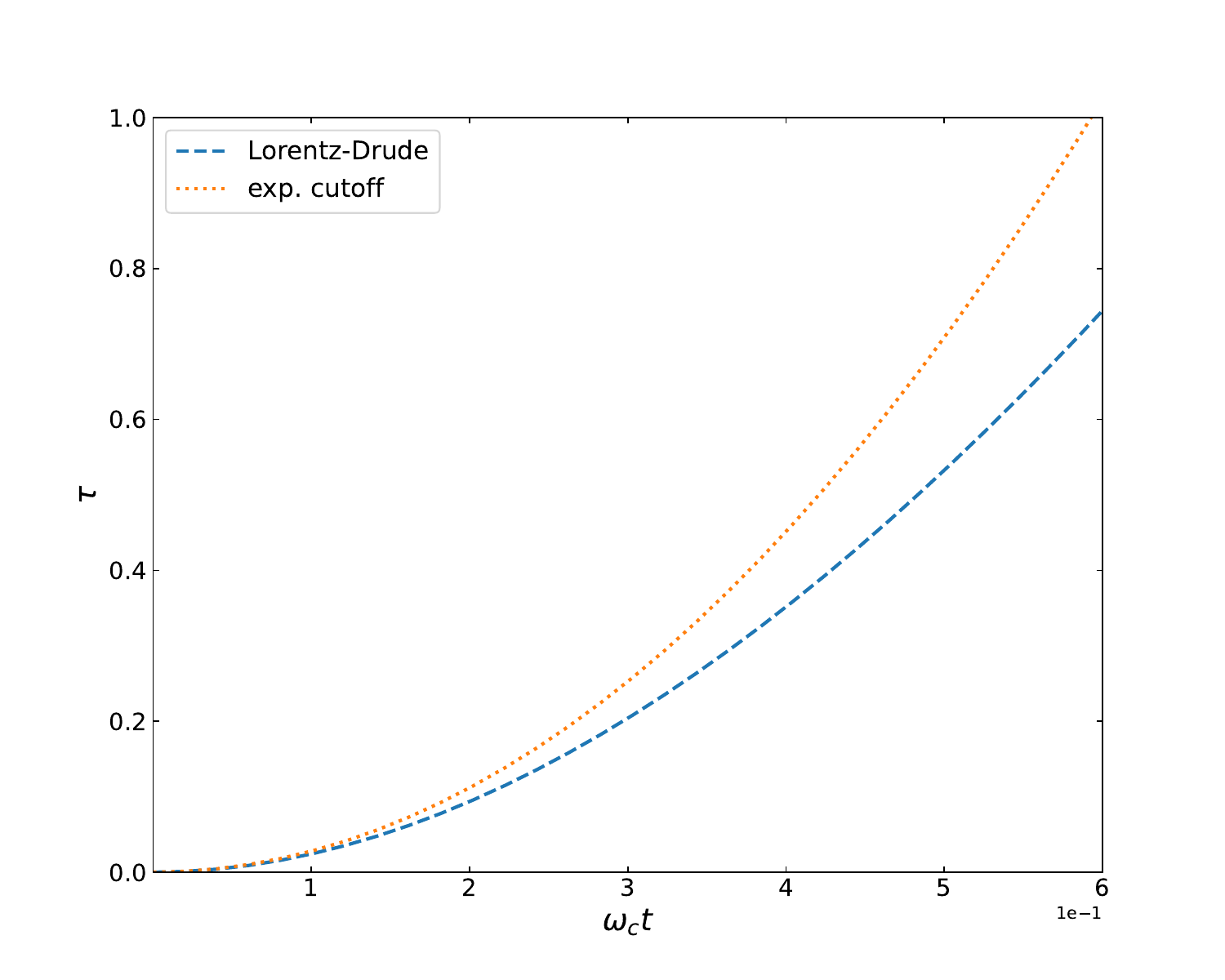}
\caption{$\tau$ for Ohmic spectral density is plotted against dimensionless time $\omega_c t$ for Lorentz-Drude and exponential cutoff. We have considered same cutoff for the both cases $\omega_c$ and same dimensionless temperature $\gamma T/\omega_c=5$.}
\end{figure}

\pagebreak

\section{II. Teleportation Kraus operator}

\subsection{A. Revisiting GKP teleportation}
The teleportation requires a two-mode Bell state:
\begin{equation}\label{phi+}
    \ket{\Phi^+_{\beta,L}}=\frac{1}{N_{\Phi_+}}\left(\ket{0_{\beta,L}}\ket{0_{\beta,L}}+\ket{1_{\beta,L}}\ket{1_{\beta,L}}\right),
\end{equation}
where $N_{\Phi_+}$ is the normalization factor. A shifted EPR measurement is then performed jointly on the input mode, which carries the logical information to be teleported, and the first mode of the entangled pair. 
The shifted (EPR) projections are defined by the two-mode maximally entangled states
\begin{equation}\label{EPR}
    \ket{EPR(\mathbf{m})}=\frac{1}{\sqrt{2\pi}}\int dr \, e^{ipr}\ket{r}_{\hat{q}_1}\ket{r+q}_{\hat{q}_2},
\end{equation}
where $\mathbf{m}=(q,p)$ are parameters that determine the measurement outcomes. These states form a complete basis for the two-mode continuous variable Hilbert space. 
Consequently, the identity operator on the two-mode space can be resolved as
\begin{equation}
    \hat{I}\otimes\hat{I} = \int d^2\mathbf{m} \; \ket{EPR(\mathbf{m})}\bra{EPR(\mathbf{m})}.
\end{equation}
For further simplicity of the notations, we denote the EPR projectors as
\begin{equation}
    \hat{\Pi}_{EPR}(\mathbf{m})= \ketbra{EPR(\mathbf{m})}.
\end{equation}
In quantum optics, the EPR measurement is implemented using linear optics. 
The two modes are first mixed on a $50{:}50$ beam splitter. 
This is followed by homodyne measurements of the canonical quadratures on the output modes where 
the position and momentum measurement outcomes are $q/\sqrt{2}$ and $p/\sqrt{2}$, respectively.
The EPR measurement effectively teleports the input state to the output port and implements a displacement depending on the measurement outcome $\mathbf{m}=(q,p)$, 
\begin{equation} 
\label{kraus_gkp}
    K(q,p) = \hat{\Pi}_{\beta}\hat{D}\left(-\mathbf{m}\right),
\end{equation}
where $\hat{\Pi}_{\beta}=\ketbra{0_{\beta}}+\ketbra{1_{\beta}}$ is the unnormalized (quasi) projector onto the approximate GKP codespace.  

The correction restores the state to the input logical state, up to a small bit-flip error arising from the finite squeezing of the GKP state or, equivalently, from the nonzero overlap between the logical states $|0_\beta\rangle$ and $|1_\beta\rangle$. The (un-normalized) teleported mode without the quadrature corrections is then given by
\begin{equation}
    \label{rho_out_without_deco}
    \boxed{\rho_{out} =  \hat{\Pi}_{\beta}\hat{D}\left(-\mathbf{m}\right) \rho_{in}\hat{D}^\dagger\left(-\mathbf{m}\right) \hat{\Pi}^\dagger_{\beta}.}
\end{equation}
It is not hard to see that due to the GKP projection $\hat{\Pi}_{\beta}$, the output state $\rho_{out}$ is in approximate-GKP subspace.

Now, we derive an integral expression equivalent to Eq.~\eqref{rho_out_without_deco}. Let $A$ be the input mode, $B$ and $C$ the entangled modes over which the two mode GKP Bell state is shared. The shared Bell state \eqref{phi+} is expressed in operator form, upto the normalization factor, as
\begin{equation}
    \begin{aligned}
        |\Phi^+_{\beta} \rangle \langle \Phi^+_{\beta}| = \int dx \, dx^\prime\, dy \, dy^\prime \, 
        \Phi^+_{\beta}(x, x^\prime; y, y^\prime)\ketbra{x, y}{x^\prime, y^\prime},
    \end{aligned}
\end{equation}
where we denote $\Phi^+_{\beta}(x,x^\prime;y,y^\prime) = \langle{x, y}|{\Phi^+_{\beta}}\rangle\langle{\Phi^+_{\beta}}|{x^\prime, y^\prime}\rangle$. The EPR projection on the input mode $A$ and first mode of the Bell pair, namely $B$, transforms the two mode position basis states as
\begin{equation}
    \hat{\Pi}_{EPR}(\mathbf{m}) \ketbra{z, x}{z^\prime, x^\prime} \hat{\Pi}_{EPR}(\mathbf{m})^\dagger = \frac{1}{2\pi}e^{ip(z^\prime-z)}\delta(z+q-x)\delta(z^\prime+q-x^\prime) \hat{\Pi}_{EPR}(\mathbf{m}).
\end{equation}
Let $\rho_{in}$ be the density operator of the input mode, then the EPR projection followed by tracing out $A$ and $B$ gives
\begin{equation} \label{rho_out_cv_unnorm}
    \rho^\prime_C = \frac{1}{2\pi}\int dx \, dx^\prime\, dy \, dy^\prime \, e^{ip(x^\prime-x)} 
        \Phi^+_{\beta}(x, x^\prime; y, y^\prime)\langle x-q|\rho_{in}|x^\prime - q \rangle\ketbra{y}{y^\prime}_C.
\end{equation}

Due to the measurement projection, the output state $\rho_{\mathrm{out}}$ is generally unnormalized in the form given by Eq.~\eqref{rho_out_cv_unnorm}, or equivalently Eq.\eqref{rho_out_without_deco}. For the moment, we need not be concerned with the normalization.
Eq.~\eqref{rho_out_cv_unnorm} can be further simplified by translation of the integral variables, $x\mapsto x+q, x^\prime\mapsto x^\prime+q$, and therefore after setting $C$ as the output mode
\begin{equation}
\label{rho_out_without_deco_trans}
\boxed{\rho_{out} = \frac{1}{2\pi}\int dx \, dx^\prime\, dy \, dy^\prime \, e^{ip(x^\prime-x)}
        \Phi^+_{\beta}(x+q, x^\prime+q; y, y^\prime)\langle x|\rho_{in}|x^\prime \rangle\ketbra{y}{y^\prime}.}
\end{equation}

The normalization factor in both the expressions for $\rho_{out}$, Eqs.~\eqref{rho_out_without_deco} and \eqref{rho_out_without_deco_trans}, is same and given by the EPR-measurement outcome probability
\begin{equation}
P(\mathbf{m}) = \tr\left[\hat{\Pi}_{EPR}(\mathbf{m})\,\left(\rho_{in}\otimes\ketbra{\Phi^+_{\beta}}\right)\right].
\end{equation}
Therefore, from Eqs.~\eqref{rho_out_without_deco} and \eqref{rho_out_without_deco_trans} we can establish a mathematical equivalence:
\begin{equation}
    \label{rho_out_equi}
    \boxed{
    \hat{\Pi}_{\beta}\hat{D}\left(-\mathbf{m}\right) (\cdot)\hat{D}^\dagger\left(-\mathbf{m}\right) \hat{\Pi}^\dagger_{\beta}
    = \frac{1}{2\pi}\int dy \, dy^\prime \,\left[\int dx \, dx^\prime\, e^{ip(x^\prime-x)}
        \Phi^+_{\beta}(x+q, x^\prime+q; y, y^\prime)\langle x|(\cdot)|x^\prime \rangle\right]\ketbra{y}{y^\prime}
    .}
\end{equation}
Here, $(\cdot)$ denotes the density operator of a bosonic mode. Next, we will use this equivalence to derive the teleportation Kraus operator for the decohered ancilla Bell pair.

\subsection{B. Teleportation Kraus operator when the ancilla Bell pair is decohered.}

The GKP Bell pair used for the teleportation undergoes pure dephasing decoherence. With Eq.\eqref{decoh_factor}, the two-mode state is now given by
\begin{equation}
\Phi^+_{\beta}\left(x,x';y,y'\right) \mapsto \Phi^+_{\beta}\left(x,x';y,y'\right)\exp\left[
        -\tau(t)((x-x')+(y-y'))^2
    \right].
\end{equation}
Therefore, with Eq.\eqref{rho_out_cv_unnorm}, it is not difficult to see that the teleportation output becomes
\begin{equation}
    \rho^\prime_{out} = \frac{1}{2\pi}\int dx \, dx^\prime\, dy \, dy^\prime \, e^{ip(x^\prime-x)} 
        \Phi^+_{\beta}(x, x^\prime; y, y^\prime) \exp\left[
        -\tau(t)((x-x')+(y-y'))^2 \right]\langle x-q|\rho_{in}|x^\prime - q \rangle\ketbra{y}{y^\prime}.
\end{equation}
With translation of the integral variables, $x\mapsto x+q, x^\prime\mapsto x^\prime+q$, 
\begin{equation} \label{rho_out_decohered_bell}
\rho_{out} = \frac{1}{2\pi}\int dx \, dx^\prime\, dy \, dy^\prime \, e^{ip(x^\prime-x)}
        \Phi^+_{\beta}(x+q, x^\prime+q; y, y^\prime) \exp\left[
        -\tau(t)((x-x')+(y-y'))^2 \right] \langle x|\rho_{in}|x^\prime \rangle\ketbra{y}{y^\prime}.
\end{equation}
The decoherence factor can be expressed in Fourier transform as,
\begin{equation} \label{deco_factor_fourier}
    \exp\left[-{\tau(t)}((x^\prime-x)+(y^\prime-y))^2\right] = \int dr \,\mathcal{G}_{0,\tau(t)}(r)\,e^{ir((x^\prime-x)+(y^\prime-y))},
\end{equation}
where
\begin{equation}
    \mathcal{G}_{0,\tau(t)}(r) = \frac{1}{2\sqrt{2\pi}{\tau(t)}}\exp(-\frac{r^2}{4\tau(t)^2})
\end{equation}
is a Gaussian normal distribution centered at zero and with standard deviation $\sqrt{2}\tau(t)$. After substituting Eq.~\eqref{deco_factor_fourier} into Eq.~\eqref{rho_out_decohered_bell},
\begin{equation}\label{rho_out_decohered_bell_1}
\begin{aligned}
\rho_{out} & = \frac{1}{2\pi}\int dr \mathcal{G}_{0,\tau(t)}(r)\int dy\, dy' e^{ir(y'-y)}\int dx \, dx^\prime\,  e^{i(p+r)(x^\prime-x)}
        \Phi^+_{\beta}(x+q, x^\prime+q; y, y^\prime)  \langle x|\rho_{in}|x^\prime \rangle\ketbra{y}{y^\prime}, \\
        & = \frac{1}{2\pi}\int dr \mathcal{G}_{0,\tau(t)}(r) \hat{D}\left(-ir\right)\left[\int dy\, dy'\int dx \, dx^\prime\,  e^{i(p+r)(x^\prime-x)}
        \Phi^+_{\beta}(x+q, x^\prime+q; y, y^\prime)  \langle x|\rho_{in}|x^\prime \rangle\ketbra{y}{y^\prime}\right]\,\hat{D}^\dagger\left(-ir\right). \\
\end{aligned}
\end{equation}
Here, we have used
\begin{equation}
    \hat{D}\left(-ir\right)\ketbra{y}{y'}\hat{D}^\dagger\left(-ir\right) = e^{ir(y'-y)}\ketbra{y}{y'}
\end{equation}
for the last step.

Now, we can use the Kraus--state equivalence relation, Eq.~\eqref{rho_out_equi}, derived for the GKP teleportation in the previous subsection 
\begin{equation} \label{rho_out_kraus_1}
    \rho_{out} = \int dr \, \mathcal{G}_{0,\tau(t)}(r) \hat{D}\left(-ir\right)\,\hat{\Pi}_{\beta}\,\hat{D}\left(-\mathbf{m}'\right) \,\rho_{in}\,\hat{D}^\dagger\left(-\mathbf{m}'\right) \hat{\Pi}^\dagger_{\beta}\,\hat{D}^\dagger\left(-ir\right).
\end{equation}
Here, $\mathbf{m}'\equiv (q, p+r)$, and therefore $\hat{D}(-\mathbf{m}')=\hat{D}(-\mathbf{m}-ir)$.
Using the Weyl relation,
\begin{equation}
    D(\alpha)D(\beta)
    =
    e^{i\,\mathrm{Im}(\alpha\beta^*)}
    D(\alpha+\beta),
\end{equation}
we can decompose the displacement
\begin{equation}
    D(-\mathbf m') = e^{-i\phi(\mathbf m,r)}D(-\mathbf m)D(-ir),
\end{equation}
where the phase $\phi(\mathbf{m}, r)=\Im{ir(q+ip)}$ disappears in operator formalism because of conjugate multiplication. With this, Eq.\eqref{rho_out_kraus_1} reads
\begin{equation}
    \rho_{out} = \int dr \, \mathcal{G}_{0,\tau(t)}(r) \hat{D}\left(-ir\right)\,\hat{\Pi}_{\beta}\,\hat{D}(-\mathbf{m})\,\hat{D}(-ir)\, \,\rho_{in}\,\hat{D}^\dagger(-ir)\,\hat{D}^\dagger(-\mathbf{m}) \,\hat{\Pi}^\dagger_{\beta}\,\hat{D}^\dagger\left(-ir\right).
\end{equation}
It is easy to see that the Kraus operator of the teleportation can, therefore, be expressed as
\begin{equation}
\boxed{\hat{\mathcal{E}}_{\mathbf{m},\tau(t)}(\cdot)
     =
    \int dr\,\mathcal{G}_{0,\tau(t)}(r)
    \mathcal{K}_{\mathbf{m}}(r)(\cdot)\mathcal{K}^\dagger_{\mathbf{m}}(r)}
\end{equation}
where,
\begin{equation}
     \hat{\mathcal{K}}_{\mathbf{m}}(r)
    =
    \hat{D}(-ir)\,
    \hat{\mathcal{T}}_\beta(\mathbf{m})\,
    \hat{D}(-ir),
\end{equation}
and 
\begin{equation}
    \hat{\mathcal T}_\beta(\mathbf m)
    =
    \hat{\Pi}_\beta\hat{D}(-\mathbf m)
\end{equation}
is the teleportation Kraus in absence of decoherence.

\end{document}